\documentclass[pre,twocolumn,aps,floats,floatfix,superscriptaddress]{revtex4}

\usepackage{graphicx}
\usepackage{amsmath, amsthm, amssymb}
\usepackage{epsfig}

\begin{document}

\title{Structure of urban movements: polycentric activity and entangled hierarchical flows}

\author{Camille Roth}
\affiliation{CAMS (CNRS/EHESS) 54, boulevard Raspail\\
  F-75006 Paris, France\\}
\affiliation{Institut des Syst\`emes Complexes de Paris-Ile de France (ISC-PIF), 57-59 rue Lhomond  , F-75005 Paris, France\\
  \texttt{\small roth@ehess.fr}}

\author{Soong Moon Kang}
\affiliation{Department of Management Science and Innovation\\
University College London (UCL), Gower Street, London WC1E 6BT, UK\\
 \texttt{\small smkang@ucl.ac.uk}}

\author{Michael Batty} \affiliation{Centre for Advanced Spatial
 Analysis (CASA)\\ University College London (UCL), 1-19 Torrington
 Place, London WC1E 7HB, UK\\
 \texttt{\small m.batty@ucl.ac.uk}}

\author{Marc Barth\'elemy}
\affiliation{CAMS (CNRS/EHESS) 54, boulevard Raspail\\
F-75006 Paris, France\\}
\affiliation{Institut de Physique Th\'eorique\\
CEA, IPhT, CNRS-URA 2306, F-91191 Gif-sur-Yvette, France\\
\texttt{\small marc.barthelemy@cea.fr}}

\date{\today} \widetext
\begin{abstract}

The spatial arrangement of urban hubs and centers and how individuals
interact with these centers is a crucial problem with many
applications ranging from urban planning to epidemiology.  We utilize
here in an unprecedented manner the large scale, real-time 'Oyster'
card database of individual person movements in the London subway to
reveal the structure and organization of the city.  We show that
patterns of intraurban movement are strongly heterogeneous in terms of
volume, but not in terms of distance travelled, and that there is a
polycentric structure composed of large flows organized around a
limited number of activity centers.  For smaller flows, the pattern of
connections becomes richer and more complex and is not strictly
hierarchical since it mixes different levels consisting of different
orders of magnitude. This new understanding can shed light on the
impact of new urban projects on the evolution of the polycentric
configuration of a city and the dense structure of its centers and it
provides an initial approach to modeling flows in an urban system.
  
\end{abstract}

\maketitle

\section{Introduction}

The structure of a large city is probably one of the most complex
spatial system that we can encounter. It is made of a large number of
diverse components connected by different transportation and
distribution networks.  In this respect, the popular conception of a
city with one center and pendular movements going in and out of the
business center is likely to be an audacious simplification of what
actually happens. The most prominent and visible effects of such
spatial organization of economic activity in large and densely
populated urban areas are characterized by severe traffic congestion,
uncontrolled urban sprawl of such cities and the strong possibilities
of rapidly spreading viruses biologial and social through the dense
underlying networks \cite{Eubank:2004,Wang,Balcan}. The mitigation of
these undesirable effects depends intrinsically on our understanding
of urban structure \cite{Batty:book}, the spatial arrangement of urban
hubs and centers, and how the individuals interact with these
centers. The dominant model of the industrial city is based on a
monocentric structure~\cite{Krugman,Wilson}, but contemporary cities
are more complex, displaying patterns of polycentricity that require a
clear typology for their understanding \cite{Kloosterman:2000}. One of
the most important features of an urban landscape is the clustering of
economic activity in many centers \cite{Anas:1998}: the idea of the
polycentric city in such terms can be traced back over one hundred
years \cite{Friedmann:1965,Geddes}, but so far no clear quantitative
definition has been proposed, apart from various methods of density
thresholding based, for example, on employment
\cite{Thurstain:2000}. In order to characterize polycentricity, we
must investigate movement data such as person flow and mobile-phone
usage \cite{Gonzalez:2008} which offers the possibility of analyzing
quantitatively various features of the spatial organization associated
with individual traffic movements. More precisely, in this study, we
analyze data for the London underground rail (`tube') system collected
from the Oyster card (an electronic ticketing system used to record
public transport passenger movements and fare tariffs within Greater
London) which enables us to infer the statistical properties of
individual movement patterns in a large urban setting.

\section{Results}

World cities \cite{Hall:1984} are among those with the most complex spatial
structure. The number, the diversity of components and their
localization warns us intuitively that these megapoles are far from
their original historical form which is invariably represented by a
simple, monocentric structure. In particular, the level of commercial
and industrial activity varies strongly from one area to another. Thus
flows of individuals can be thought as good proxies for the activity
of an area and to this end we first checked that the flows at
different stations correlate positively with other activity indicators
such as counts of employees and the employee density. This shows that
indicators of a different nature and on different time scales, which
are also widely regarded as measures of polycentricity in large
cities, are also consistent with movement data recorded over much
shorter time scales.

The main results that we will discuss in this section are that (i)
flows are generally of a local nature (ii) they are also
organized/aggregated around polycenters and (iii) the examination and
decomposition of these flows lead to the description of entangled
hierarchies, and (iv) hence one likely structure describing this large
metropolitan area is based on polycentrism. This perspective thus
draws new insights from data that has become available from electronic
sources that have so far not been utilised in analyzing the urban
spatial structure and in this sense, are unprecedented in the field.

To get a preliminary grasp on the data, we observe that the flow
distribution (normalized histogram of flows of individuals) is fitted
by a power law with exponent $\approx 1.3$ which indicates that there
is strong heterogeneity of individuals' movements in this city (for
this distribution, the ratio of the two first moments has a large
value $\langle w^2\rangle/\langle w\rangle^2\simeq 15.0$, which
confirms this strong heterogeneity)--- see Figure~$1$.  
\begin{figure}[!h]
\centering
\begin{tabular}{c}
\includegraphics[angle=0,scale=1.0]{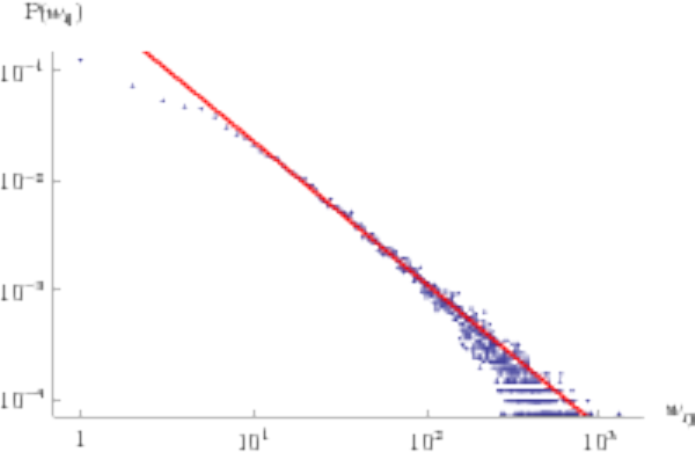}
\end{tabular}
\caption{ {\bf Flow distribution}. Loglog plot of the
histogram of the number of trips between two stations of the 
tube system. The line is a power law fit with exponent $\approx 1.3$.}
\label{fig:1}
\end{figure} 
Broad distribution of flows have already been observed at the
inter-urban level \cite{deMontis:2007}, but it is the first time that
we observe this empirically at an intra-urban level showing that, in
agreement with other studies (for Madrid \cite{Gutierrez:2006} and for
Portland, Oregon \cite{Eubank:2004}), the movement patterns in large
cities exhibit an heterogeneous organization of flows.

Spatial separation is another primary feature of movement and we show
in Figure~$2a$ the raw distribution of rides occurring between two
stations at a given distance. This distribution can be fitted by a
negative binomial law rather than a broad law such as the Levy flights
suggested in \cite{Gonzalez:2008,Brockmann:2006}.
\begin{figure*}
\centering
\begin{tabular}{c}
\includegraphics[angle=0,scale=0.8]{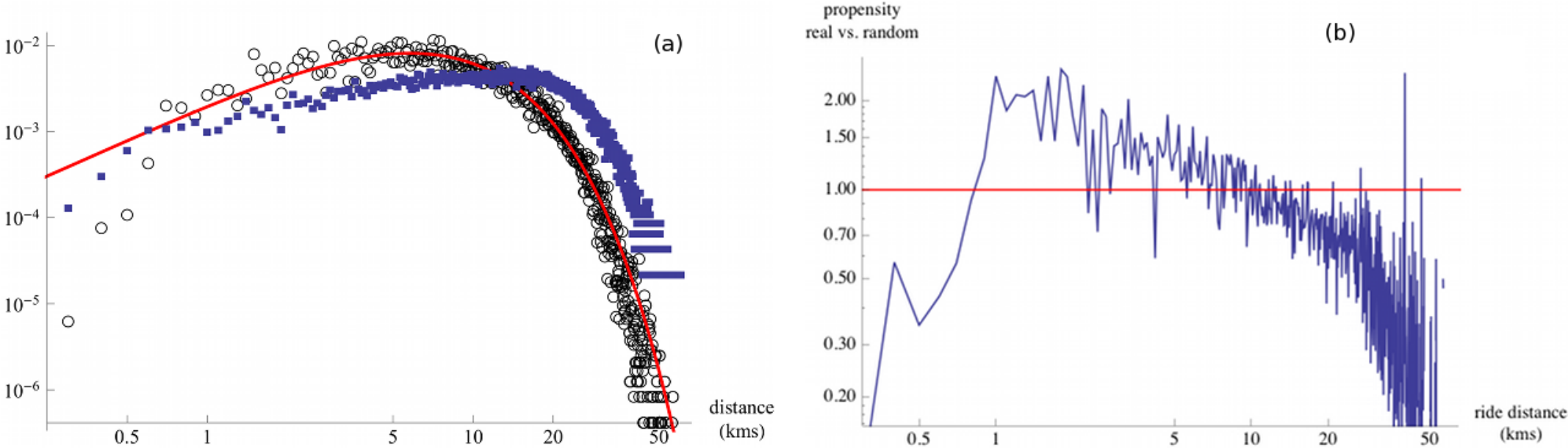}
\end{tabular}
\caption{ {\bf Ride distance distribution and propensity.} (a)
Superimposition of the distance distribution of rides (circles) and of
the distance distribution between stations (squares). The distribution
of the observed rides can be fitted by a negative binomial law of
parameters $r=2.61$ and $p=0.0273$, corresponding to a mean
$\mu=9.28$kms and standard deviation $\sigma=5.83$kms (solid
line). This distribution is not a broad law (such as a Levy flight for
example), in contrast to previous findings using indirect measures of
movement \cite{Gonzalez:2008,Brockmann:2006}. (b) Ride distance
propensity. Propensity of achieving a ride at a given distance with
respect to a null-model of randomized rides.}
\label{fig:2}
\end{figure*} 
While this graph exhibits actual commuting patterns, it does not tell
us much about commuter behavior, all other things being equal.
Indeed, the geographical constraints are important and the distance
distribution between stations (shown superimposed in Figure $2a$)
could be a major factor in the ride distribution. Also, the particular
flow distribution over the network is likely to bias the ride distance
distribution: rides corresponding to two stations, which have
respectively a large outflow and inflow, should be more likely, hence
the distance between these two stations is likely to be
overrepresented in the previous distribution. This bias relates to how
much agents prefer to use the underground to achieve rides at a given
distance. In order to estimate the part governed by the individuals
behavior, we use a null-model for randomizing rides in such a way that
total outflows and total inflows at each station are conserved while
actual ride extremities are reshuffled (see Appendices). Put differently,
the random null-model corresponds to a flow matrix that should
normally occur given particular out- and inflows at stations,
irrespective of agent's preferences. Dividing the real-world values by
the random flow matrix (averaged over $100$ random simulations) gives
the propensity (see Appendices) which is an estimate of how much the real
data deviates from a random setting. Results are described in
Figure~$2b$. We observe that rides covering a distance of around $1$
to $3$kms are twice as likely. The propensity continuously falls to
$0$ for longer rides, and is significantly less than one for rides of
less than $1$km. Above a distance of $10$kms, the propensity is less
than one indicating that individuals are less inclined to use the
subway for longer distances. Hence, all other things being equal,
people are less inclined to take the tube for rides not covering this
sort of `typical' distance.

In addition to being strongly heterogenous, rides are therefore to
some extent essentially local. At a more aggregated level, and in
order to infer the city structure at a larger scale, we can study the
distribution of incoming (or outgoing) flows for a given station. We
show in the Figure~$3$ the rank-ordered total flows (Zipf plots) for
the morning peak hours on a lin-log graph displaying an exponential
decay (Flows for evening peak hours ($5$pm-$8$pm) reveal a roughly
inverse pattern, \hbox{i.e.} the total outflow is concentrated on a
few centers, and similarly but less markedly, the same occurs for
total inflows).
\begin{figure}[!h]
\centering
\begin{tabular}{c}
\includegraphics[angle=0,scale=1.0]{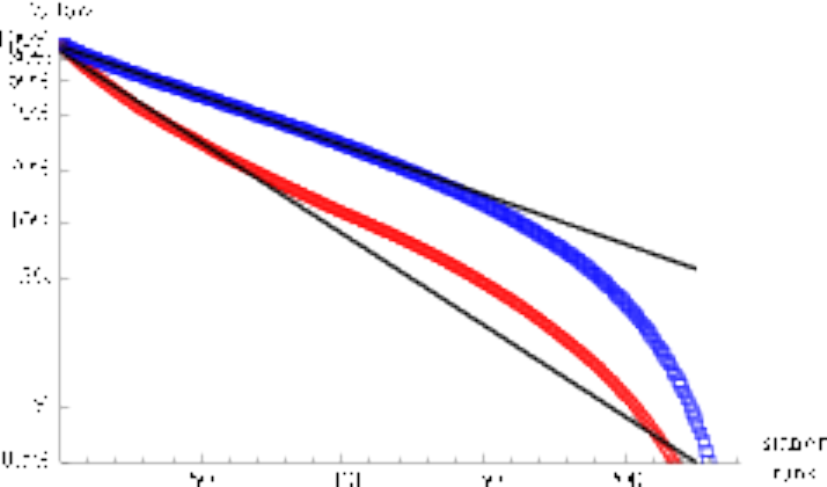}
\end{tabular}
\caption{ {\bf Total flow distributions.} Zipf plot for the total
  inflows {\em (red circles, below)} and total outflows {\em (blue
    squares, above)} for morning peak hours (7am-10am). The inflow $I$
  (outflow $O$) of a station $j$ ($i$) is defined as
  $I(j)=\sum_iw_{ij}$ ($O(i)=\sum_jw_{ij}$). The straight lines are
  exponential fits of the form $e^{-r/r_0}$ with $1/r_0^{in}\simeq
  2.27\cdot 10^{-2}$ for the inflow and $1/r_0^{out}\simeq
  1.40\cdot10^{-2}$ for the outflow.}
\label{fig:3}
\end{figure} 
The exponential decay of these plots demonstrate that most of the
total flows are concentrated on a few stations. Indeed, an exponential
decay of the form $e^{-r/r_0}$, where $r$ is the rank, is a signature
of the existence of a scale $r_0$. In this case, the exponential fit
shows that the number of important inflow stations is of order $n\sim
r_0^{in}\sim 45$ and larger for outflow stations. During the morning
peak hours, essentially, stations that generate a large inflow have a
smaller outflow, and vice-versa. Also, rides are
statistically balanced over the entire day, which suggests that rides
are essentially round trips. From this analysis, we
can conclude that the activity is concentrated in a small number of
centers dispersed over the city. Using the exponential distribution of
flows, we can then define multiple centers acting as sources or sinks
depending on the time of day. 

To examine further this polycentric structure, we will aggregate
different stations if their inflow is large and they are spatially
close to one another. Various clustering methods could be used and we
choose one of the simplest described in the appendices. This
clustering yields a hierarchical, descending decomposition of inflows
with respect to an increasing share of the total inflow in the
network. We summarize the results of this process in the dendrogram
shown in Figure~$4$.
\begin{figure*}
\centering
\begin{tabular}{c}
\includegraphics[angle=0,scale=0.9]{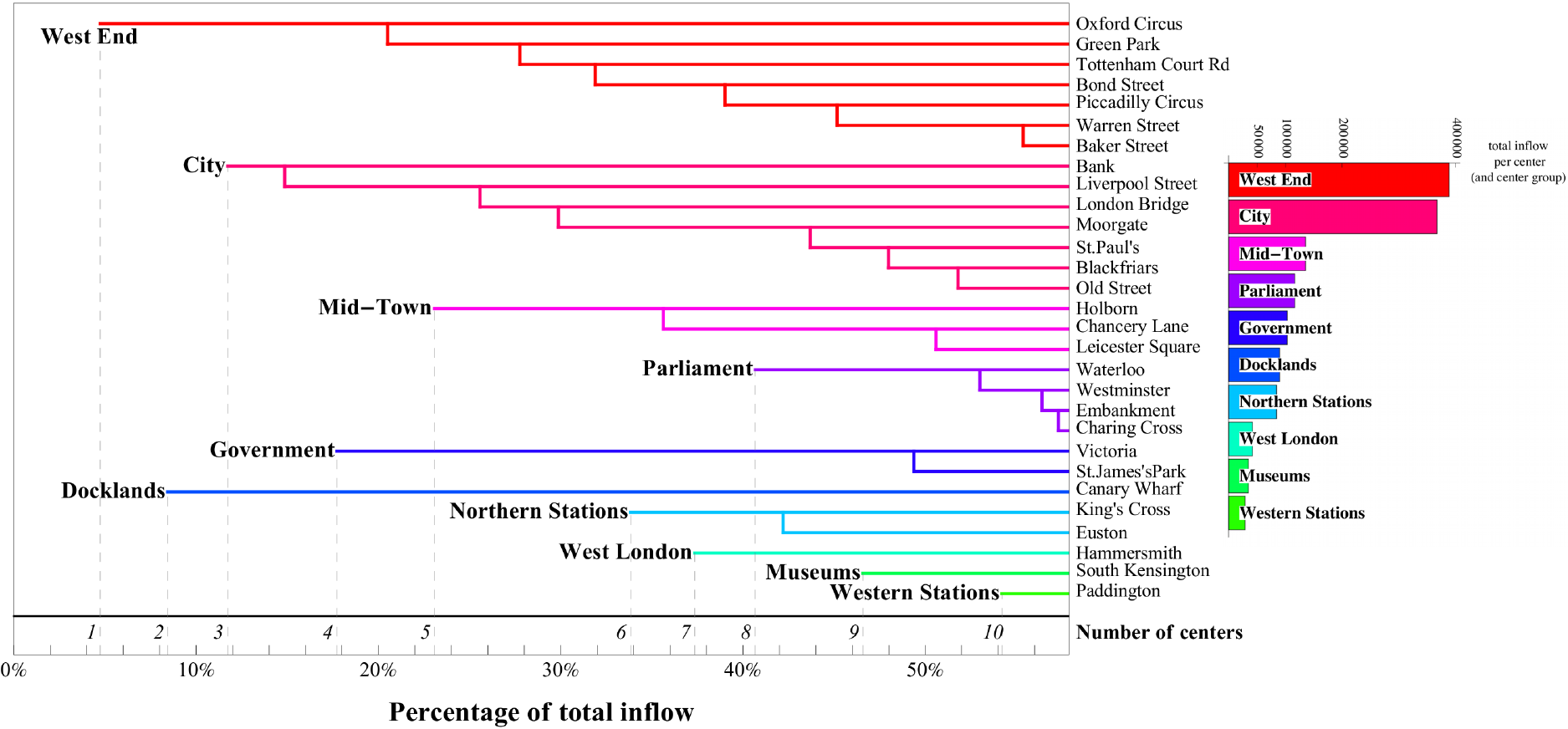}
\end{tabular}
\caption{ {\bf Hierarchical organization of the
    activity: Polycenters.} Breakdown of centers in terms of
  underlying stations and inflows.  We gather stations by descending
  order of total inflow and we aggregate the stations to centers when
  taking into account more and more stations. In this process, all
  stations within $1,500$ meters of an already-defined center are
  aggregated to this main center. This yields the dendrogram shown
  here which highlights the hierarchical nature of the polycentric
  organization of this urban system. The bold names to the left of the
  aggregates --- such as ``\emph{West End}'' for the group of stations
  around Oxford Circus --- are used throughout the paper as convenient
  labels to denote the polycenters.}
\label{fig:4}
\end{figure*} 
This dendrogram highlights the hierarchical organization of urban
polycentricity. The number of centers is not an absolute quantity, but
depends on an observation scale as measured here by the percentage of
inflow. As we consider higher percentages of the total inflow, more
centers are taken into account, which leads to centers as an aggregate
of multiple sub-centers with smaller inflows. In other words, this is
equivalent to saying that at large spatial scales, we observe one
large center corresponding to the whole city, and when we decrease the
scale of observation, multiple centers appear, which are themselves
composed of smaller centers. This hierarchical nature is crucial and
indicates that we cannot define a center by applying a threshold rule
(e.g., an area is a center if the population or employment density is
larger than some threshold \cite{Thurstain:2000}), but that it can
only be defined according to a given scale.

We represent the ten most important polycenters defined in the
dendrogram of Figure~$4$, and show the corresponding propensity to
anisotropy comparing actual flows with the null model defined above
(see the appendices). This comparison shows that the actual flows are in
general very different from what is obtained using the random null
model. We study the relative orientation of the incoming flow
(normalized by its corresponding quantity given by the null model) and
picture it by eight-segment compasses, which we show in Figure~$5$ on
the central and inner London underground map. 
\begin{figure*}
\centering
\begin{tabular}{c}
\includegraphics[angle=0,scale=0.5]{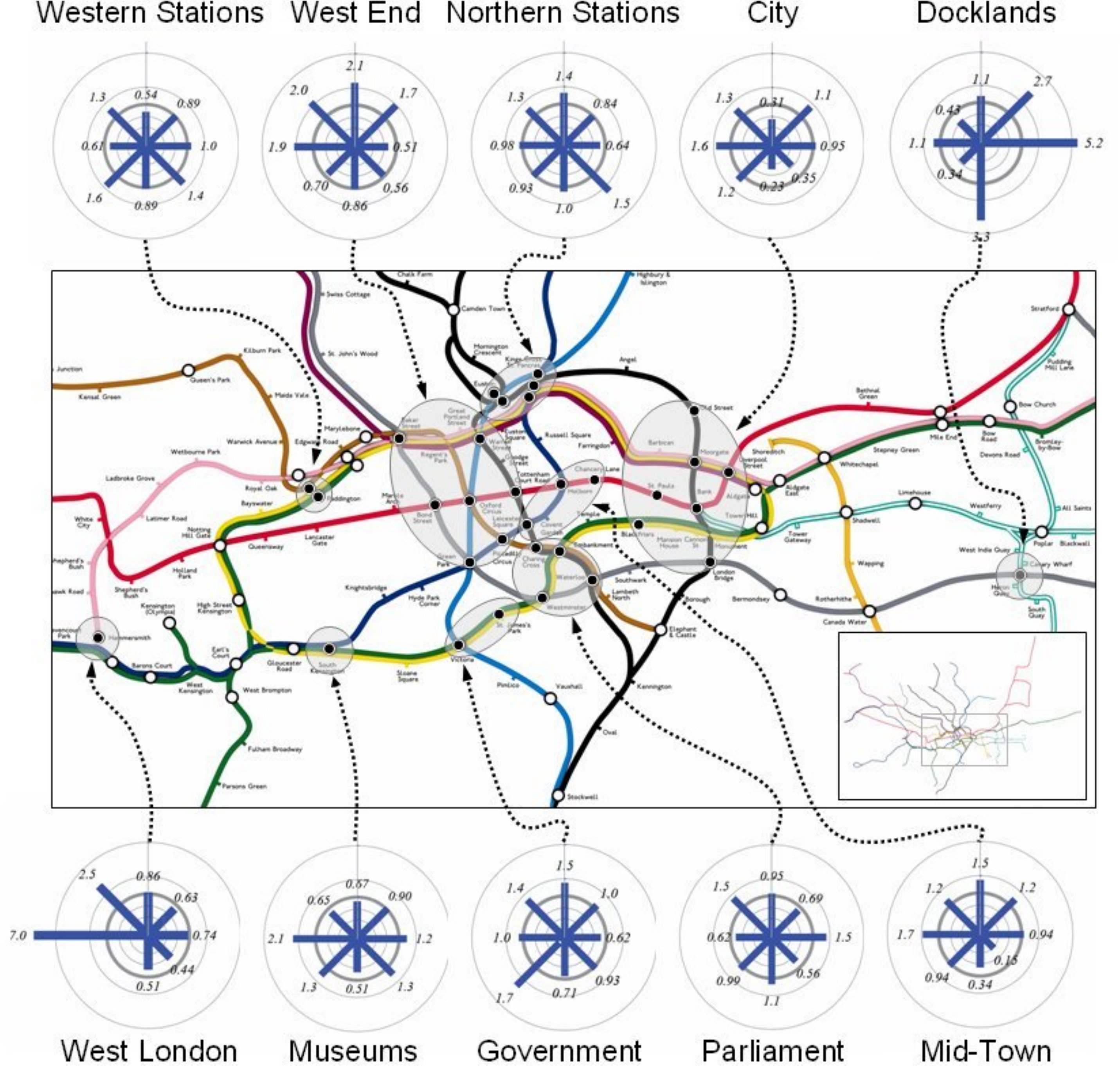}
\end{tabular}
\caption{ {\bf The London subway (tube) system: polycenters and basins
    of attraction.} In the inset, we show the entire tube network
  while in the main figure, we zoom in on the central part of
  London. We represent the ten most important polycenters defined in
  the dendrogram of Figure~$3$, and show the corresponding propensity
  to anisotropy comparing actual flows with the null model defined in
  the text. A propensity of $1$ means that there is no deviation in a
  given direction with respect to the null model. Circles correspond
  to various levels of identical propensity values: the thicker circle
  in the middle corresponds to $1$, inner circles correspond to
  propensities of $0.2$ and $0.5$, and outer circles to $2$ and
  $5$. The anisotropy is essentially in opposite directions from the
  center, thus showing a strong bias towards the suburbs for
  peripheral centers essentially, rather than for central
  centers. Moreover, most stations control their own regions and seem
  to have their own distinctive basins of attraction.}
\label{fig:5}
\end{figure*} 
The absence of any bias would give a fully isotropic compass with all
segments of radius equal to one (propensity equal to $1$). The
anisotropy is essentially in opposite directions from the center, thus
showing a strong bias towards the suburbs essentially for peripheral
rather than for central centers.


We now examine how the flows are distributed into and outside centers,
focusing on the morning peak hours. We first aggregate the flows by
centers by computing the total flow incoming to a certain center $C$:
\begin{equation}
w_{iC}=\sum_{j\in C}w_{ij}
\end{equation}
In this aggregated view, we thus represent movements by a directed
network where flows go from single stations (the “sources”) to
centers, which are groups of stations.

We then rank all flows $w_{iC}$ in a decreasing order, thereby
focusing on paths of decreasing importance as if we were detailing a
map starting with highways, then concentrating on roads, and then on
streets. We consider the $N$ most important flows such that the
corresponding sum of flows is a given percentage $W$ of the total flow
in the network. For example, if we consider the flows up to $W=20\%$
of the total flow, we obtain the structure that we show in Figure~$6$
(it should be noted that we kept the `station-to-center' flows such
that they represent $20\%$ of the total flow, which is different from
keeping the most important station-to-station flows such as it is done
for the Figure~$4$ precisely in order to define those `centers'. We
thus cannot directly compare these Figures $4$ and $6$).
\begin{figure}[!h]
\centering
\begin{tabular}{c}
\includegraphics[angle=0,scale=0.17]{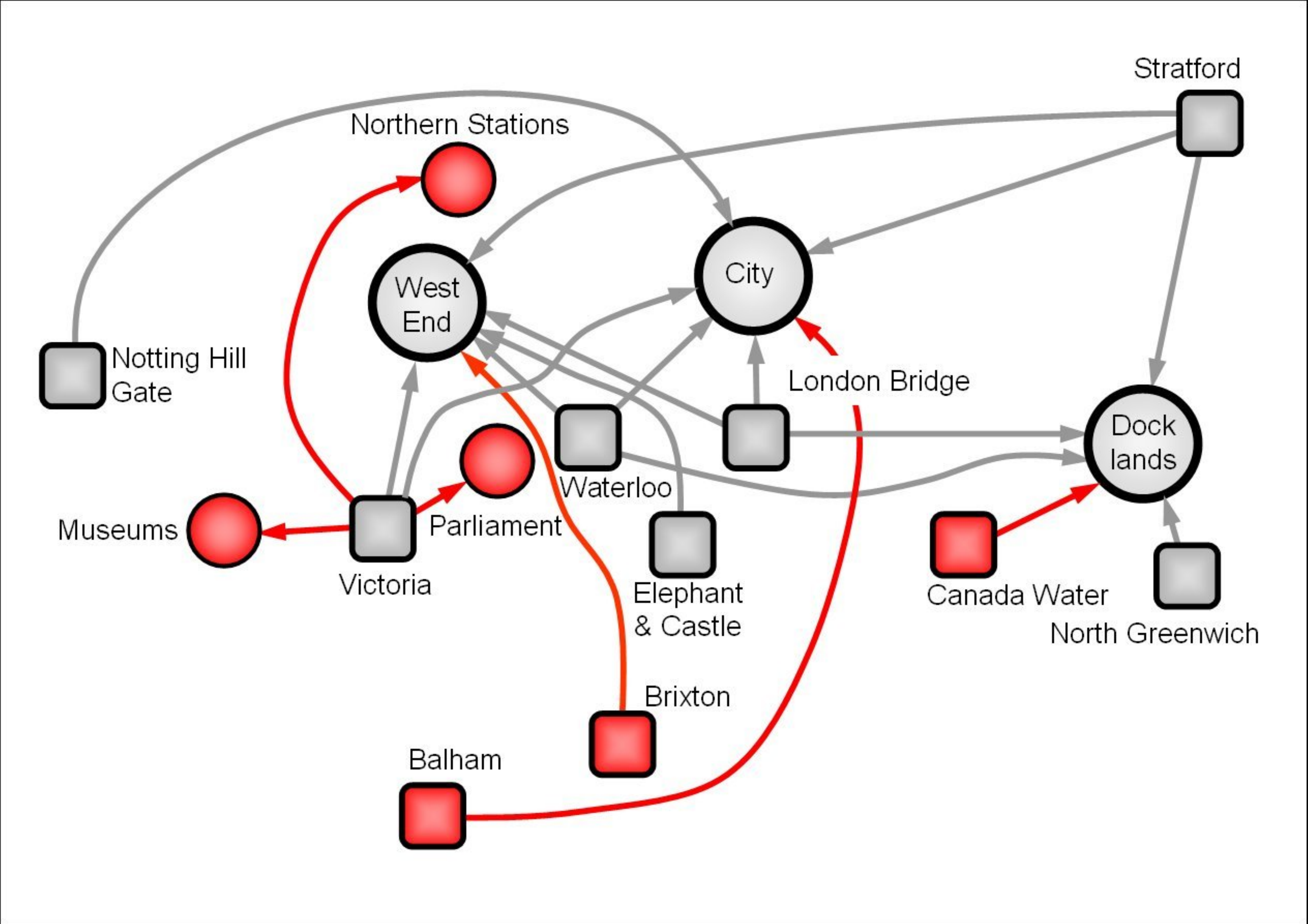}
\end{tabular}
\caption{ {\bf Structure of flows at $20\%$ and $40\%$ of the total
    flow.} When considering the most important flows from stations to
  centers such their sum represents $20\%$ of the total flow in the
  network, we observe sources (represented as squares) with outdegree
  $k_{out}=3$ such as London Bridge, Stratford, or Waterloo connecting
  to three different centers (represented as circles), as well as
  sources with $k_{out}=2$ (eg. Victoria) and $k_{out}=1$
  (eg. Elephant and Castle). We also show how the pattern of flows is
  constructed iteratively when we go to larger fraction of the total
  flow (from $20\%$ shown in black to $40\%$ shown in red). We
  represent in red the new sources, centers and connections. The new
  sources connect to the older centers (eg. \emph{West End},
  \emph{City}, etc) and the existing sources (eg. \emph{Victoria})
  connect to new centers (eg.  \emph{Northern stations},
  \emph{Museums}, and \emph{Parliament}).}
\label{fig:6}
\end{figure} 

At this scale, it is clear that we have three main centers and sources
(with various outdegree values), which mostly correspond to intermodal
rail-subway connections. Adding more links, we reach a fraction $W =
40\%$ of the total flow and we then investigate smaller flows at a
finer scale. We see that we have new sources appearing at this level
and new connections from sources that were present at $W=20\%$.

We can summarize this result with the graph shown in Figure~$7$ where
we divide the centers into three groups according to their inflow
(decreasing from first Group I to the last Group III). 
\begin{figure}[!h]
\centering
\begin{tabular}{c}
\includegraphics[angle=0,scale=1.0]{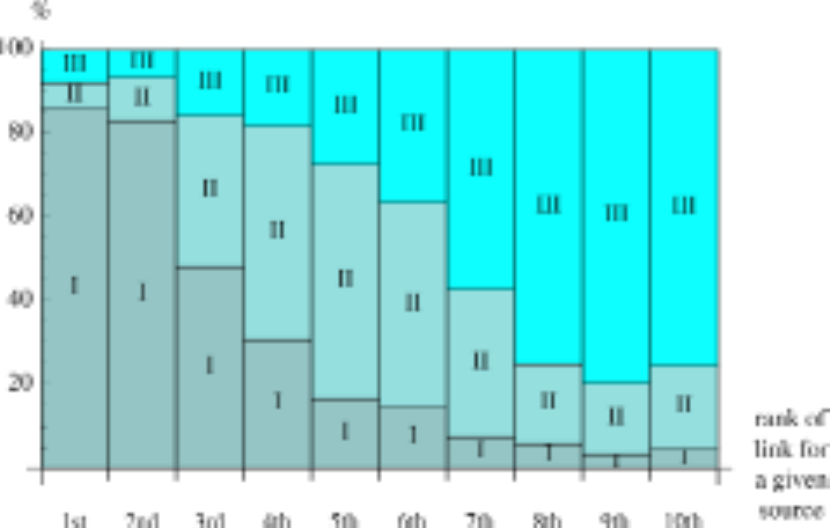}
\end{tabular}
\caption{ {\bf Most important links.} Proportion of links going from
  sources to centers of a certain group (I, II, III), considering links of
  decreasing importance for each given source, when raising $W$ (from
  the first link appearing, at left, to the last link, at right).}
\label{fig:7}
\end{figure} 
In other words (see Figure~$4$), Group I gathers centers with the most
important \emph{total} inflow namely the \emph{West End}, \emph{City}
and \emph{Mid-town}. Group II gathers the next three centers
\emph{Parliament}, \emph{Government} and \emph{Docklands} while Group
III gathers the other centers such as the \emph{Northern stations},
\emph{West London}, \emph{Museums} and the \emph{Western
  stations}. This figure shows that for more than $80\%$ of the
sources, the most important link (ie. the $1$st link) connects to a
center of Group~I. Conversely for more than $80\%$ of the sources, the
least important link (ie. $10$th link) goes to a center of
Group~III. The flow structure thus follows an original yet simple
pattern when we explore smaller and smaller weights.

We can quantify in a more precise way how the structure of flows
evolves when we investigate smaller flows by exploring the list of
flows $w_{iC}$ in decreasing order and by introducing the transition
matrix $T$, which describes how the outdegree of a source varies with
increasing $W$ (see Appendices). When we explore
smaller flows, the analysis of the T-matrix shows that the pattern of
connections from sources to centers becomes richer and more complex,
but can nonetheless be described by the simple iterative process
described above: the most important link of a source goes to the most
important centers, the second most important link connects to the
second most important centers, and so on. It is interesting to note
that even if the organization of flows follows a simple iterative
scheme, it leads to a complex and rich structure, which is not
strictly hierarchical since it mixes different levels of flows
consisting of different orders of magnitude. In addition, the fact
that the most important flows always connect to the same center
naturally leads to the question of efficiency and congestion in such a
system. In this respect, London appears as a `natural' city as opposed
to an `artificial' city for which flows would be constructed according
to an optimized, hierarchical schema \cite{Alexander:1966,Batty:2006}.

\section{Discussion}

World cities such as London have tended to defy understanding hitherto
because simple hierarchical subdivision has ignored the fact that
their polycentricity subsumes a pattern of nested urban
movements. Using the Oyster data we can identify multiple centers in
London, then describe the traffic flowing into these centers as a
simple hierarchic decomposition of multiple flows at various scales.
In other words, these movements define a series of subcenters at
different levels where the complex pattern of flows can be unpacked
using our simple iterative scheme based on the representation of ever
finer scales defined by smaller weights. Casual observation suggests
that this kind of complexity might apply to other world cities such as
Paris, New York or Tokyo where spatial structure tends to reveal
patterns of polycentricity considerably more intricate than cities
lower down the city size hierarchy. Our approach needs to be extended
of course to other modes of travel, which will complement and enrich
the analysis of polycentricity. The Oyster card is already used on
buses and has just expanded beyond the tube system to cover other
modes of travel such as surface rail in Greater London. With GPS
traffic systems monitoring, in time, all such movements will be
captured, extending our ability to understand and plan for the
complexity that defines the contemporary city.

\section{Acknowledgments}

The Oyster card data was collected by Transport for London (TfL), and
we are grateful for their permission to use it in this paper. We
also thank Cecilia Mascolo for access to TfL and the Oyster card data,
and Andrew Hudson-Smith for providing the London underground map.

\section{Appendices}

\subsection{Data}

Our analysis of individual movements is based on a dataset describing
the entire underground service between $31$ March $2008$ and $6$ April
$2008$ encompassing a total of $11.22$ million trips from $2.03$
million individual Oyster card IDs. For each trip, the data includes
the origin and destination for individual passengers as well as the
corresponding time of the trip. We stress that the data we obtained
from Transport for London (TfL) is completely anonymized without any
possibility of trace back to individuals.  Besides, we only have
individual trajectories, but not the history of the trajectories over
a long period of time which then could provide the capability of
identifying individuals from the electoral register and business
directories. From this dataset, we build the (origin/destination) flow
matrix $w_{ij}$, which gathers the aggregated number of rides leaving
a station $i$ to a station $j$ over a given period of time. The
analysis of these flow matrices in several time intervals for every
single day in the dataset shows that the commuting patterns during
weekdays present a regular and distinctive pattern in contrast to
travel at weekends. As a result, we focus our study on the commuting
patterns during weekdays.

\subsection{The null model, propensity, and anisotropy}

{\it The null model}

The subway infrastructure imposes a certain number of physical
constraints which can affect various distributions. This is for
example the case of the ride distribution where rides between two
stations with large outflow and inflow, respectively, are likely to be
over-represented. As such the ride distribution could simply be a
result of the peculiar subway spatial structure. In order to eliminate
this type of biases, we use for comparison a null-model constructed in
the following way. We randomize rides in a such a way that the total
outflow and total inflow of each station is conserved while actual
ride extremities are reshuffled. This model is basically a
configuration model \cite{molloy,newman} which preserves the total
number of incoming and outgoing links for each station and where each
link corresponds to a given ride. Put differently, the random setting
corresponds to a flow matrix (obtained here by an average over $100$
random simulations) that should normally occur given particular out-
and in-flow heterogeneity at stations, irrespective of agent
preferences.

{\it The ride propensity}

We can then divide the real values of flows $w_{ij}$ by the random flow matrix
which yields an estimate of how much the real data deviates from a
random setting (at fixed inflow-outflow constraints). For the ride
distribution we then obtain the {\it ride propensity} $R$ shown in Figure~$1b$
\begin{equation}
R(d)=\frac{1}{N(d)}\sum_{ij/d(i,j)=d}\frac{w_{ij}}{w^{nm}_{ij}}
\end{equation}
where $w^{nm}_{ij}$ is the number of individuals going from $i$ to $j$
in the null model, $d(i,j)$ represents the distance on the network
between $i$ and $j$, and where $N(d)$ is the number of pairs of nodes
at distance $d$. This propensity gives an estimate of how much the
real data deviates from a random flow assignment with the same
geographical and flow constraints. In other words, when the propensity
is equal to one the observed flows are entirely due to the
geographical and flow structure of the network. Conversely when the
propensity is smaller or larger than $1$, the flows reflect
non-uniform preferences for rides of certain distance.

{\it The anisotropy propensity}

We used the null model in order to extract the part due to the
behavior of the commuters in their ride distribution. We can also
study the relative orientation of the incoming flow normalized by its
corresponding quantity given by the null model which gives the
anisotropy $A$ due to the commuters behavior
\begin{equation}
A(\theta)=\frac{1}{N(\theta)}\sum_{ij/\widehat{iOj}=\theta}\frac{w_{ij}}{w^{nm}_{ij}}
\end{equation}
where $\theta$ is a particular direction (we binned the angle in eight
equal intervals so to represent an eight-segment compass) and where
the sum is over the $N(\theta)$ nodes $i$ and $j$ such that the angle
of $i-j$ is given by $\theta$. The absence of any bias would give a
fully isotropic compass with all segments of radius equal to one
(anisotropy propensity equal to $1$).

\subsection{Identifying the polycenters}

Clustering methods for point in spaces has been the subject of many
studies and are used in many different fields. In particular, in
computational biology and bioinformatics, clustering is used to build
group of genes with related expression patterns. Many different
methods were developped and the most common ones are hierarchical
clustering methods (such as those based on K-means and their
derivatives, see for example \cite{Clustering}). Here, we are in a
slightly different position. The stations are clearly located in space
and thus Euclidean distance appears as the natural distance measure (a
necessary ingredient for clustering methods). Yet these stations are
also characterized by their inflow.  For this reason, the usual
methods are not directly applicable and we thus adopted the simplest
clustering method which we describe as follows. We first gather
stations by descending order of total inflow, thereby defining centers
of decreasing importance. In order to account for geographical
proximity of groups of stations, indicating subsets of distinct
stations belonging to a single geographical center, we aggregate all
stations within a distance $r_c$ of an already-defined center. In this
way we systematically increase the total flow associated with these
centers and we continue this process until we capture a large
percentage of the total flow. We thus chose to stop at $60$ percent of
the total flow in order to avoid to include too many details and too
much noise.

We varied the value of $r_c$ from $1$ to $2$ kms and observed that our
results were stable. This stability probably comes from the fact that
the inter-distance station is of order $1.2$kms for London in $2008$
and corresponds to some psychological threshold above which
individuals prefer to take the subway if they can choose. The results
discussed above are obtained with $r_c=1500$ meters.

\subsection{The T matrix}

We face here a difficult problem: we have a complete weighted directed
network featuring flows from stations to centers, and the goal
is to extract some meaningful information. We started with the
analysis of the dominant flows and we would like to understand how the
flows are structured when we explore smaller values. In order to do
this, we introduce a `transition 'matrix $T$ which characterizes
quantitatively the changes in the flow structure when we explore the
list of flows $w_{iC}$ going from a station $i$ to a center $C$ in
decreasing order of importance. In what follows, when we talk of
`total flow at $W$', we mean that we consider only the most important
flows $w_{iC}$ so that we reach a total fraction $W$ of the total flow
on the whole network of station-to-center flows.  When the total
flow goes from $W$ to $W+\delta W$, the elements $t_{ij}$ of $T$
represent the number of sources with outdegree $i$ at $W$ and with
outdegree $j$ 
at $W+\delta W$. Note that $i$ starts at $i=0$ while $j$ starts at
$j=1$ (\hbox{i.e.} $T$ only denotes sources that have a strictly
positive outdegree at $W+\delta W$).
 
As an example, when we go from $W=20\%$ to $W+\Delta W=40\%$, the $T$ matrix
is
\begin{equation}
T=
\left(
\begin{array}{ccccc}
37 & 12 & 1 & 0 & 0 \\
4  & 9  & 4 & 1 & 0\\
0  & 4  & 2 & 1 & 2\\
0  & 0  & 0 & 2 & 1\\
0  & 0  & 0 & 0 & 0\\
0  & 0  & 0 & 0 & 0
\end{array}
\right)
\label{eq:tt}
\end{equation}

The matrix $T$ is composed of three parts (see Figure~$8$). 
\begin{figure}[!h]
\centering
\begin{tabular}{c}
\includegraphics[angle=0,scale=0.5]{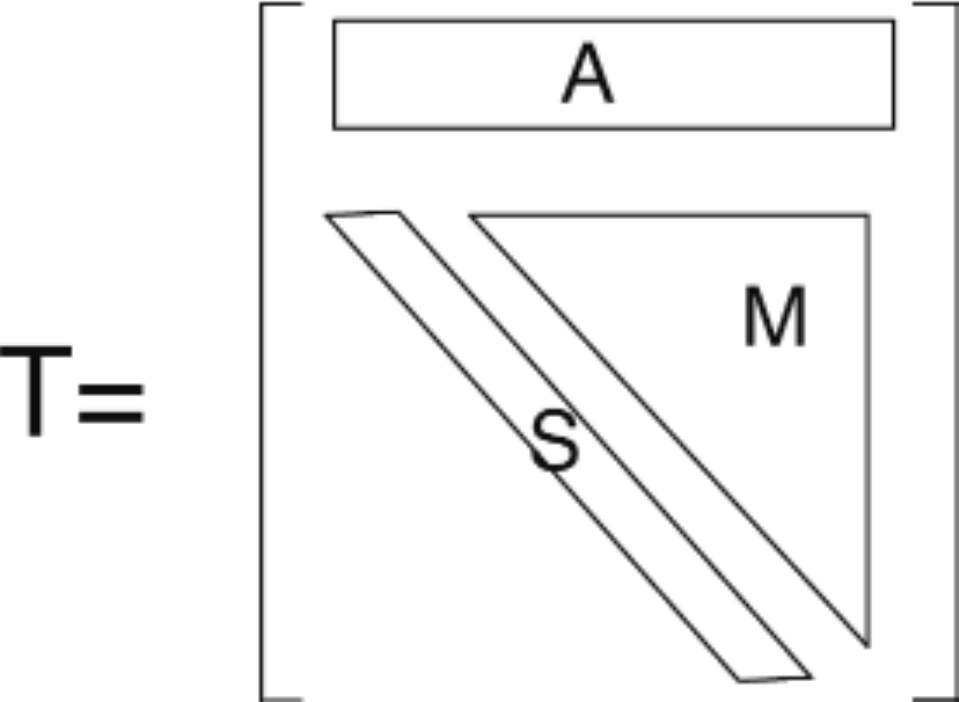}
\end{tabular}
\caption{ {\bf Transition matrix.} Typical form of the outdegree
  transition matrix $t_{ij}$, consisting essentially of a row vector
  ($A$, inexistent sources before the transition) and an upper
  triangular matrix (made of a diagonal $S$ of sources having the same
  out-degree after the transition, and a submatrix $M$ of sources
  whose out-degree increases after the transition).}
\label{fig:8}
\end{figure} 
The first part, $A$, consists of new sources appearing when we
increase the total flow, and corresponds to the first line of $t_{ij}$
where $i=0$. The second part, $S$, consists of sources where the
outdegree stays invariant when we change from $W$ to $W +\delta W$
(i.e., the diagonal $t_{ii}$). The third part, $M$, consists of
sources that were already present at the $W$ level and the outdegree
changes during the process from $W$ to $W + \delta W$ (i.e., the upper
triangle $t_{ij}$ where $j>i$). We can compute the number of sources
in each of these types and plot them. A proper $T$ matrix is a
$(N+1)\times N$ matrix (in Eq.~\ref{eq:tt}, $N=5$), as the $T$ matrix
is made of a row vector ($A$) and an upper triangular matrix ($S$, $M$
and the zeros) because a source that feeds $n$ centers cannot become a
source feeding $n'<n$ centers when transitioning to a larger
inflow-cut $W+dW$. The row vector $A$ indicates sources that were not
feeding centers before, and now feed some centers, \hbox{i.e.},
sources that were non-existent for a lower inflow-cut, hence the extra
initial row represented by vector $A$.  Thus, `$37$' means that after
the transition (at the new inflow-cut), there are $37$ new sources
feeding one center, $12$ new sources feeding two, $1$ new source
feeding three. The `$9$' on the second row means that $9$ sources that
used to feed one center, now feed two, and so on. The row $A$ is thus
given by
\begin{equation}
A=
\left(
\begin{array}{ccccc}
37 & 12 & 1 & 0 & 0
\end{array}
\right)
\end{equation}
and the diagonal is
\begin{equation}
S=
\left(
\begin{array}{ccccc}
4 & 4 & 0 & 0 & 0
\end{array}
\right)
\end{equation}
The upper triangular matrix $M$ is given by
\begin{equation}
M=
\left(
\begin{array}{cccc}
9 & 4 & 1 & 0\\
0 & 2 & 1 & 2\\
0 & 0 & 2 & 1\\
0 & 0 & 0 & 0\\
\end{array}
\right)
\end{equation}

In the case of the transition $20\%\rightarrow 40\%$, the major
phenomenon is the appearance of new sources ($37$ in this case)
followed by sources feeding new centers.

Figure~$9$a shows the number of new sources ($A$ in the matrix $T$) and
the sources that change type ($S$). 
\begin{figure}
\centering
\begin{tabular}{c}
\includegraphics[angle=0,scale=1.0]{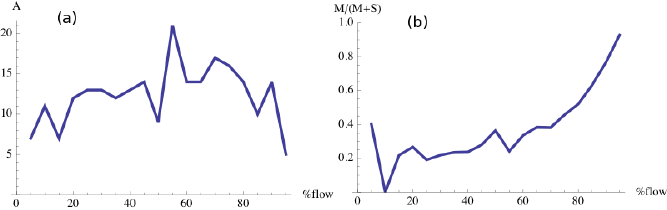}
\end{tabular}
\caption{ {\bf Evolution of the number of sources and their type.} (a) Number
  of new sources ($A$) versus the total flow $W$. (b) Fraction of existing
  sources whose type is changing ($M$) when the total flow varies from $W$
  to $W+\delta W$. Here $\delta W=5\%$.}
\label{fig:9}
\end{figure} 
We observe that there is a continuous addition of new sources along
with connections to new and old centers. Besides, for a total flow
less than $50\%$, there is a relatively stable proportion of sources
(about $20\%$) whose outdegree varies when $W$
increases. 
When we zoom into finer scales (i.e., larger values of the
total flow $W$), new sources appear and connect preferentially to the
existing largest centers, while the existing sources connect to the
new centers through secondary connections. This yields two types of
connection only. The first type goes from new sources to old centers,
and the second type from old sources to new centers.











\end{document}